\documentclass[%
 reprint,
superscriptaddress,
 amsmath,amssymb,
 aps,
 pre,
floatfix,
]{revtex4-2}

\usepackage[textsize=tiny]{todonotes}

\usepackage{graphicx}
\usepackage{dcolumn}
\usepackage{bm}
\usepackage{mathtools}
\usepackage{hyperref}

\renewcommand{\eqref}[1]{Eq.~(\ref{#1})}
\newcommand{\eqsref}[1]{Eqs.~(\ref{#1})}
\newcommand{\figref}[1]{Fig.~\ref{#1}}
\newcommand{\figsref}[1]{Figs.~\ref{#1}}
\newcommand{\secref}[1]{Sec.~\ref{#1}}
\newcommand{\secsref}[1]{Secs.~\ref{#1}}
\newcommand{\appref}[1]{Appendix~\ref{#1}}

\begin{document}

\title{Interplay between self-assembly and phase separation in a polymer-complex model}

\author{Tianhao Li}%
\affiliation{Department of Chemistry, Princeton University, Princeton, NJ 08544, USA}

\author{W. Benjamin Rogers}
\affiliation{Martin A. Fisher School of Physics, Brandeis University}

\author{William M. Jacobs}
\email{wjacobs@princeton.edu}
\affiliation{Department of Chemistry, Princeton University, Princeton, NJ 08544, USA}

\date{\today}

\begin{abstract}
  We present a theoretical model for predicting the phase behavior of polymer solutions in which phase separation competes with oligomerization.
  Specifically, we consider scenarios in which the assembly of polymer chains into stoichiometric complexes prevents the chains from phase-separating via attractive polymer--polymer interactions.
  Combining statistical associating fluid theory with a two-state description of self-assembly, we find that this model exhibits rich phase behavior, including re-entrance, and we show how system-specific phase diagrams can be derived graphically.
  Importantly, we discuss why these phase diagrams can resemble---and yet are qualitatively distinct from---phase diagrams of polymer solutions with lower critical solution temperatures.
\end{abstract}

\maketitle

\section{Introduction}

Predicting the phase behavior of biopolymer systems is challenging due to the diversity of conformational and oligomeric states that biopolymers can populate~\cite{shin2017liquid,pappu2023phase}.
In this context, oligomers refer to small molecular clusters that form due to attractive interactions among biomacromolecules, including proteins~\cite{li2012phase,lan2022quantitative} and nucleic acids~\cite{jain2017rna,nguyen2022condensates}.
The interplay between oligomerization and phase separation is complex and can give rise to qualitatively different behaviors in different scenarios.
In some cases, phase separation correlates with oligomer stability, either because phase separation is enhanced by the presence of oligomers~\cite{li2012phase,marzahn2016higher,conicella2020tdp,kar2022phase,lan2022quantitative,he2023phase} or because phase separation promotes oligomer formation in condensed phases~\cite{shin2017liquid,weber2019spatial}.
Yet in other cases, phase separation appears to compete with oligomer formation, meaning that the presence of stable oligomers tends to inhibit demixing into condensed biopolymer-rich phases~\cite{seim2022dilute}.

Here we explore this latter scenario---competition between the self-assembly of oligomers and phase separation---using a thermodynamically consistent mean-field model.
We consider polymer solutions in which multiple polymer chains can assemble into a \textit{stoichiometric complex}, which is an oligomer comprising a precise number of chains in a well-defined geometry~\cite{marsh2015structure}.
A theoretical challenge is that this scenario cannot be described in terms of pairwise interactions alone, since the assembly of discrete stoichiometric complexes implies that spatial correlations among multiple polymer chains cannot be ignored~\cite{wertheim1984fluids,jacobs2014phase}.
We therefore propose a coarse-graining strategy in which a two-state model of stoichiometric complex self-assembly is coupled to a mean-field model of interacting, yet spatially uncorrelated, polymer chains within the framework of statistical associating fluid theory (SAFT)~\cite{chapman1989saft}.

This theoretical approach treats self-assembly and phase separation in a self-consistent framework~\cite{reinhardt2011re,bauermann2022chemical}, allowing us to predict the effects of self-assembly on the equilibrium phase behavior of associating polymer solutions~\cite{semenov1998thermoreversible}.
In the present article, we systematically develop this theoretical approach and study the implications of the resulting model under wide-ranging scenarios.
In a closely related article~\cite{hegde2023competition}, we separately provide validation of our theoretical predictions in a particular DNA-based experimental system, which utilizes the programmability of Watson--Crick base-pairing~\cite{santalucia2004thermodynamics} to tune both the thermodynamic stability of a designed stoichiometric complex and the phase behavior of the DNA oligonucleotides.
Taken together, the results of these studies demonstrate the validity of the assumptions employed in our theoretical approach and confirm the quantitative accuracy of our model's predictions.

The key theoretical results of our approach are presented in this paper as follows.
In \secsref{sec:model} and \ref{sec:coexistence}, we describe the mean-field model and our numerical strategy for calculating its phase behavior in various parameter regimes.
Then in \secsref{sec:conserved-volume} and \ref{sec:nonconserved-volume}, we investigate two scenarios in which the excluded volume of a stoichiometric complex is either the same as or greater than the total excluded volume of its constituent chains.
Both of these scenarios exhibit re-entrant phase behavior, in which a constant-concentration polymer solution transitions from one phase to two coexisting phases and back to a single phase as a single control parameter is tuned monotonically.
This control parameter represents an experimentally accessible quantity such as, but not limited to, temperature, ionic strength, pH, or a co-solvent concentration.
Importantly, although many features of the phase behavior resemble qualitative aspects of phase diagrams with a \textit{lower critical solution temperature (LCST)}, we find that competition between self-assembly and phase separation does not result in a critical point of this type.
Instead, we find that phase coexistence can terminate at a first-order transition due to the competition between self-assembly and phase separation.
Our approach shows how qualitatively different phase diagrams can arise by altering the dependence of key dimensionless parameters on experimentally controllable variables.
A more thorough discussion of the implications of these results is provided in \secref{sec:discussion}.

\section{Theoretical Model and Results}

\subsection{Two-state self-assembly and phase separation}
\label{sec:model}

\begin{figure}[t]
\includegraphics[width=\columnwidth]{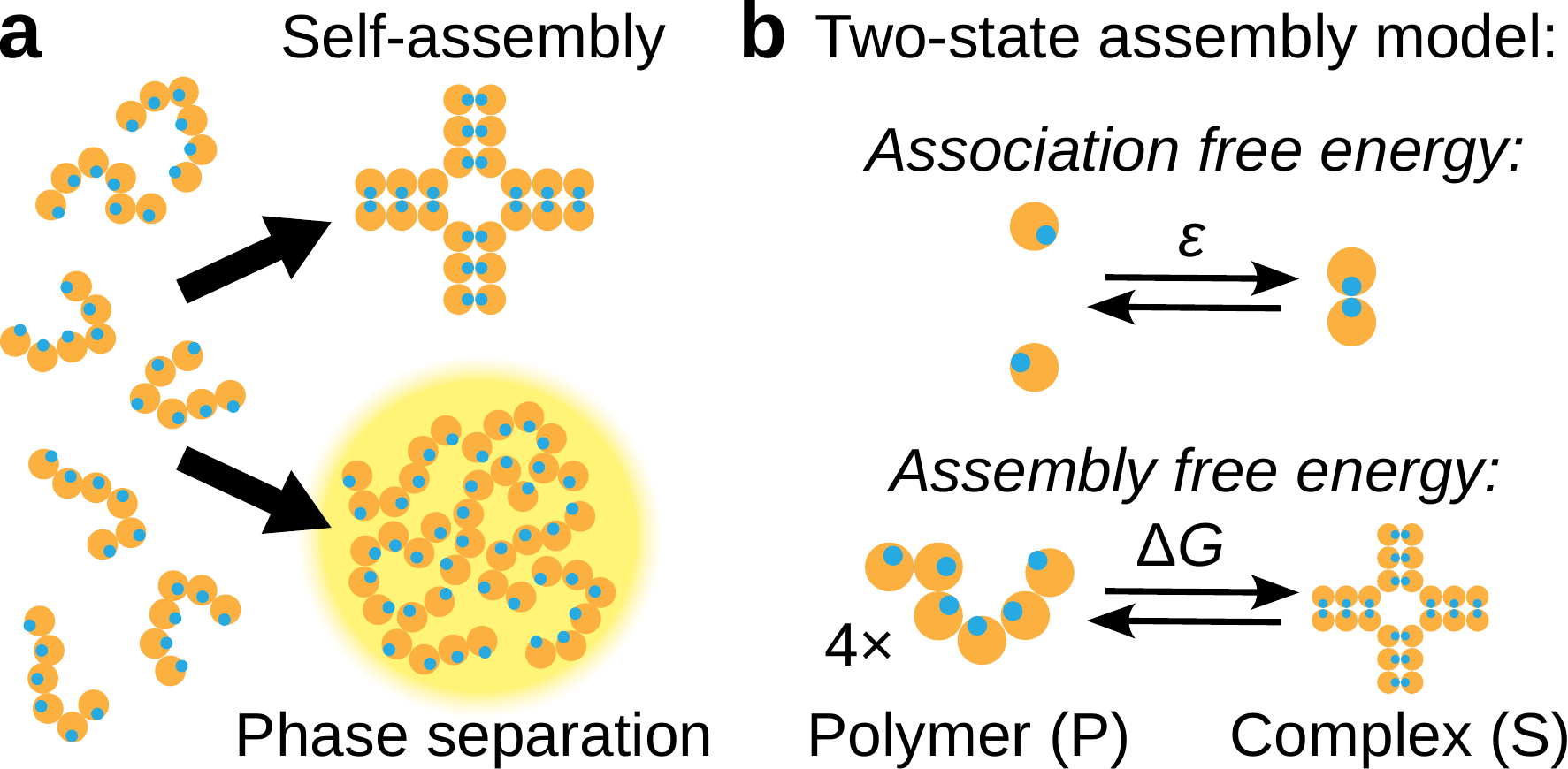} 
\caption{\textbf{Schematic of the theoretical model.}
  (a)~Biopolymers can either assemble into stoichiometric complexes or form a condensed phase via phase separation.  Polymers are represented as freely jointed chains of blobs (yellow circles), each of which contains a single binding site (blue dots).  Here we consider stoichiometric complexes consisting of $n=4$ polymers.
  (b)~The model involves two independent free-energy parameters, $\epsilon$ and $\Delta G$, which govern the pairwise association between binding sites on blobs and the assembly of chains into stoichiometric complexes, respectively.  Associative interactions between blobs are treated using statistical associating fluid theory (SAFT), while the two-state model of stoichiometric complex assembly is treated as an equilibrium chemical reaction between the polymer (P) and complex (S) states.}
  \label{fig:model}
\end{figure}

We first develop a mean-field model of coupled self-assembly and liquid--liquid phase separation (\figref{fig:model}a).
We consider a polymer solution consisting of a single polymeric species in an implicit solvent.
Polymers are coarse-grained into chains of $N_{\text{p}}$ ``blobs,'' each of which represents a Kuhn segment~\cite{rubinstein2003polymer} of the polymer with excluded volume $v_0$.
This coarse-graining allows us to treat polymers as freely jointed chains, meaning that the relative orientations of adjacent blobs are uncorrelated.
The total excluded volume of a free polymer is $v_{\text{p}} = N_{\text{p}} v_0$.
To account for ``specific'' short-ranged attractive interactions, such as hybridization of DNA oligomers~\cite{santalucia2004thermodynamics} or screened electrostatic interactions between charged sidechains of polypeptides~\cite{brangwynne2015polymer}, we introduce $N_{\text{p}}$ distinct binding sites on each polymer.
Each blob can accommodate a single coarse-grained binding site within this framework, since the orientations of monomers within a Kuhn-segment blob are highly correlated and thus not independent of one another~\cite{rubinstein2003polymer}.
Specific interactions between pairs of binding sites therefore represent coarse-grained, effective interactions between two Kuhn-length groups of monomers.

This modeling approach relies on three key assumptions.
First, we assume that each binding site can participate in at most one specific attractive interaction at a time.
Second, we assume that the assembly of $n$ polymers into a stoichiometric complex is highly cooperative, such that partially assembled intermediates involving fewer than $n$ polymers are vanishingly rare at equilibrium (i.e., ``two-state'' self-assembly).
Third, we assume that all binding sites engage in specific interactions within a stoichiometric complex in the assembled state.
This final assumption can be relaxed, leaving additional binding sites on assembled complexes available for interactions with binding sites either on other complexes or on free polymers~\cite{hegde2023competition}, but we do not consider this possibility in the present work.

Taken together, these assumptions imply that self-assembly and phase separation are mutually exclusive (\figref{fig:model}a): Associative interactions between available binding sites are required to drive phase separation, and yet these binding sites are completely saturated by the formation of a complete stoichiometric complex.
We note that, in practice, satisfying the second, ``two-state'' assumption typically requires a stoichiometric mixture of $n$ different polymers with distinct binding sites, such that each polymer species is used exactly once in constructing a unique stoichiometric complex~\cite{biffi2013phase,biffi2015equilibrium}.
The model that we describe here treats all polymers as being indistinguishable for simplicity, which means that the specific interactions between free polymers represent effective interactions that average over all possible pairs of binding sites.

We implement these assumptions within a mean-field theory by combining statistical associating fluid theory (SAFT)~\cite{chapman1989saft} with a variation of the Flory--Huggins model of a polymer solution~\cite{rubinstein2003polymer}.
In accordance with the second key assumption of two-state self-assembly, chains can either exist in the free polymer state (P) or as part of a stoichiometric complex (S) consisting of $n$ chains.
The excluded volume of a complex is $v_{\text{s}}$, which we relate to the excluded volume of a polymer blob via $N_{\text{s}} \equiv v_{\text{s}} / v_0$ for notational convenience.
Importantly, because our partitioning of chains into either P and S states allows us to treat the configuration of an assembled complex independently from that of individual polymers, it is not necessarily the case that $v_{\text{s}} = n v_{\text{p}}$.
We shall consider both scenarios, where the excluded volume is either conserved or not conserved, in \secsref{sec:conserved-volume} and \ref{sec:nonconserved-volume}, respectively.

The free-energy density, $f$, comprises four terms,
\begin{equation}
  f = f^{\text{id}} + f^{\text{ev}} + f^{\text{assoc}} + f^{\text{int}},
\end{equation}
accounting for the ideal (id), excluded-volume (ev), associative (assoc), and internal (int) contributions, respectively.
Each term has units of energy per volume.
The ideal contribution depends on the concentrations (i.e., number densities) of polymers in both the P and S states,
\begin{equation}
  \label{eq:fid}
  \beta f^{\text{id}} v_0 = \frac{\phi_{\text{p}}}{N_{\text{p}}} (\ln \phi_{\text{p}} - 1) + \frac{\phi_{\text{s}}}{N_{\text{s}}} (\ln \phi_{\text{s}} - 1),
\end{equation}
where $\phi_i \equiv v_i \rho_i$ is the volume fraction and $\rho_i$ is the concentration (or number density) of state $i \in \{\text{P},\text{S}\}$.
To avoid discussing explicit temperature dependence at this point, we use the inverse temperature $\beta \equiv 1/k_{\text{B}} T$, where $k_{\text{B}}$ is the Boltzmann constant and $T$ is the absolute temperature, to express all free energies as dimensionless quantities.
We employ a mean-field formula for the excluded volume contribution that accounts for the fact that assembled complexes are essentially rigid bodies, whereas free polymers are freely jointed chains,
\begin{align}
  \label{eq:fev}
  \beta f^{\text{ev}} v_0 &= \frac{\phi_{\text{s}}}{N_{\text{s}}} + \frac{1 - \phi_{\text{s}}}{N_{\text{s}}} \ln (1 - \phi_{\text{s}}) + \frac{\phi_{\text{p}}}{N_{\text{p}}}\left[1 - \ln (1 - \phi_{\text{s}})\right] \nonumber \\
  &\qquad + \phi_0 \ln \left( \frac{\phi_0}{1 - \phi_{\text{s}}} \right),
\end{align}
where $\phi_0 \equiv 1 - \phi_{\text{s}} - \phi_{\text{p}}$ is the volume fraction not occupied by either free polymers or stoichiometric complexes.
A derivation of this formula is presented in \appref{app:EFH}, and a comparison to the Flory--Huggins model is provided in \appref{app:B2}.

Following the two-state assumption, the associative contribution only accounts for interactions among binding sites on P-state polymers.
The free-energy density is given by Wertheim's thermodynamic perturbation theory~\cite{wertheim1984fluids}, taking the P/S mixture without attractive interactions [i.e., \eqsref{eq:fid} and \ref{eq:fev})] as the reference state~\cite{jacobs2014phase},
\begin{equation}
  \label{eq:fassoc}
  \beta f^{\text{assoc}} v_0 = \phi_{\text{p}} \left( \ln X_{\text{p}} - \frac{X_{\text{p}}}{2} + \frac{1}{2} \right),
\end{equation}
where $X_{\text{p}}$ is the fraction of binding sites on P-state polymers that are not associated.
$X_{\text{p}}$ is determined by the chemical equilibrium expression
\begin{equation}
  \label{eq:Xassoc}
  X_{\text{p}} + \phi_{\text{p}} e^{-\beta \epsilon} X_{\text{p}}^2 = 1,
\end{equation}
where $\epsilon$ is the association free energy between a pair of binding sites (\figref{fig:model}b).
\eqref{eq:fassoc} can be derived by minimizing the Helmholtz free energy of a mixture of reactive binding sites; \eqref{eq:Xassoc} follows directly from the derivation of $f^{\text{assoc}}$.
We refer the reader to Ref.~\cite{michelsen2001physical} for an elegant derivation and a thorough discussion.
Multiplying \eqref{eq:Xassoc} by $\phi_{\text{p}}$, we can easily identify this chemical equilibrium condition as a mass-action expression relating the volume fraction of blobs that are not associated, $X_{\text{p}}\phi_{\text{p}}$, and the volume fraction of blobs that are associated, $(1 - X_{\text{p}})\phi_{\text{p}}$, subject to an equilibrium constant $\exp(-\beta\epsilon)$.
Association does not occur in the limit $\beta\epsilon \rightarrow \infty$.
Importantly, \eqsref{eq:fassoc} and (\ref{eq:Xassoc}) assume that the association states of different binding sites are uncorrelated.
This requirement is consistent with our coarse-grained treatment of association, in which we assume that each Kuhn-segment blob accommodates a single binding site.

By contrast with the associative interactions, the internal contribution to the free-energy density only depends on the concentration of S-state complexes,
\begin{equation}
  \label{eq:fint}
  \beta f^{\text{int}} v_0 = \frac{\phi_{\text{s}}}{N_{\text{s}}} \left[\beta \Delta G + n(1 - N_{\text{p}})\right],
\end{equation}
where $\Delta G$ represents the free-energy change (excluding translational entropy) associated with the self-assembly ``reaction'' $n\text{P} \rightleftharpoons \text{S}$ (\figref{fig:model}b).
This reaction is at equilibrium when
\begin{equation}
  \label{eq:PSeq}
  n\mu_{\text{p}} = \mu_{\text{s}},
\end{equation}
where the P and S-state chemical potentials are $\mu_{\text{p}} \equiv \partial f / \partial \rho_{\text{p}}$ and $\mu_{\text{s}} \equiv \partial f / \partial \rho_{\text{s}}$, respectively.
Intuitively, \eqref{eq:fint} says that the contribution to the free energy due to polymer--polymer interactions within an S-state complex is proportional to the concentration of S-state complexes, $\rho_{\text{s}} = \phi_{\text{s}} / N_{\text{s}} v_0$.
We motivate the particular definition of $\beta\Delta G$ established by \eqref{eq:fint} by discussing a limiting case at the end of this section.

This treatment allows us to study the consequences of stoichiometric complex self-assembly within a mean-field theory.
In reality, both self-assembly and association among free polymers are driven by the same physical interactions among binding sites (\figref{fig:model}a).
These interactions among binding sites can be assumed to be essentially pairwise in nature.
However, the probability that a specific pair of binding sites is associated is highly correlated with the binding status of proximal binding sites on the same chains when self-assembly takes place.
Such correlations are challenging to capture within mean-field models via thermodynamic perturbation theory.
By contrast, including a discrete S state via the two-state approximation and the equilibrium condition \eqref{eq:PSeq} allows us to account for these correlations, while still treating the associative interactions between binding sites on P-state polymers as uncorrelated in \eqsref{eq:fassoc} and (\ref{eq:Xassoc}).

Nonetheless, the common physical origin of polymer association and self-assembly implies that the association free energy, $\epsilon$, and the assembly free energy, $\Delta G$, are coupled parameters that must change in tandem when a control parameter is tuned.
To be precise, $\Delta G$ depends on both the sum total of the $n N_{\text{p}} / 2$ binding interactions with strength $\epsilon$ within a stoichiometric complex, as well as entropic considerations related to the geometry and flexibility of the stoichiometric complex.
In \secsref{sec:conserved-volume} and \ref{sec:nonconserved-volume}, we show how system-specific relationships between $\epsilon$ and $\Delta G$ can be used to derive a phase diagram as a function of a single control parameter.

Before proceeding, it is instructive to consider the behavior of this model in three limiting scenarios.
If $\beta\epsilon$ is finite and $\beta\Delta G \rightarrow \infty$, then self-assembly cannot occur, and the model reduces to that of an associative polymer solution in which the polymers obey the statistics of freely jointed chains~\cite{rubinstein2003polymer,semenov1998thermoreversible}.
This polymer solution has a critical point at a dimensionless associative free energy $(\beta\epsilon)_{\text{c}}$ and polymer volume fraction $\phi_{\text{c}}$.
This critical point is commonly described as an \textit{upper critical solution temperature (UCST)}, since phase separation only occurs at temperatures below $T_{\text{c}} = \epsilon k_{\text{B}}^{-1}(\beta\epsilon)_{\text{c}}^{-1}$ if $\epsilon$ is taken to be a temperature-independent constant~\cite{rubinstein2003polymer}.
Alternatively, if $\beta\epsilon$ is finite and $\beta\Delta G \rightarrow -\infty$, then the solution is a fluid of S-state complexes in implicit solvent.
The model reduces to a regular solution~\cite{porter2021phase} with excluded volume interactions only in this case, since we have assumed that there are no associative interactions among completely assembled complexes.
Lastly, if $\beta\epsilon \rightarrow \infty$ and $\beta \Delta G$ is finite, then there are no associative interactions among free polymers, and the model reduces to a concentration-dependent description of two-state self-assembly.
In this case, \eqref{eq:PSeq} simplifies to an expression of chemical equilibrium for a reaction involving ideal gases in the dilute limit, $\phi_{\text{p}} \ll 1$ and $\phi_{\text{s}} \ll 1$,
\begin{equation}
  \frac{\phi_{\text{s}}}{\phi_{\text{p}}^n} = e^{-\beta\Delta G}.
\end{equation}
This limiting scenario justifies our expression for the internal free energy in \eqref{eq:fint}.

\subsection{Phase-coexistence calculations and master phase diagrams}
\label{sec:coexistence}

\begin{figure}[htb!]
\includegraphics[width=\columnwidth]{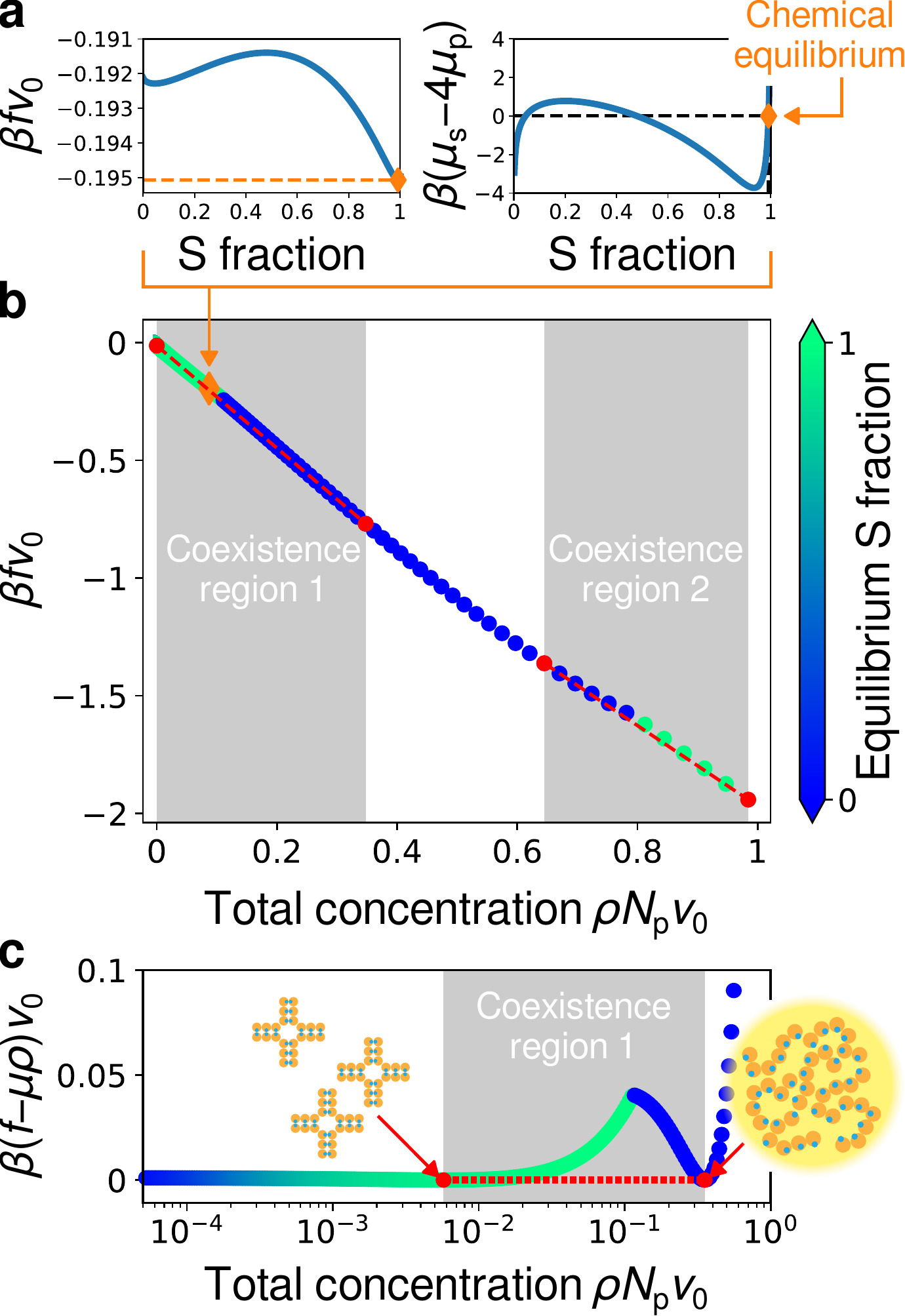} \vskip-1ex
\caption{\textbf{Schematic of phase-coexistence calculations.}
  (a)~For a given chain concentration $\rho$, we first calculate the equilibrium composition by minimizing the free energy with respect to the stoichiometric complex (S) fraction (left).  This calculation satisfies the condition of chemical equilibrium, $\mu_{\text{s}} = n \mu_{\text{p}}$, where $n=4$ in this example (right).  The equilibrium free energy is indicated by the orange diamond ($\rho N_{\text{p}} v_0 = 0.093$) assuming $N_{\text{p}} = 6$, $N_{\text{s}} = 24$, $\beta\epsilon = -4.025$, and $\beta\Delta G = -27$ (corresponding to point d in \figref{fig:conserved-volume}a).
  (b)~Performing chemical equilibrium calculations at various total concentrations results in a free-energy profile.  Points on this profile are colored according to the equilibrium fraction of chains that assemble into stoichiometric complexes.  Common-tangent constructions (red dashed lines) are then used to identify coexistence regions (shaded regions).  Red points indicate the concentrations and free energies of the coexisting phases.  In this example, there are two distinct coexistence regions at different ranges of the total concentration.
  (c)~The common-tangent construction ensures that the coexisting phases have the same chemical potential, $\mu \equiv \partial f / \partial \rho$, and the same grand potential, $f - \mu\rho$.  As an illustrative example, we show the grand potential at the coexistence chemical potential of the first coexistence region defined in panel~\textbf{b}.  The equilibrium compositions indicate that a dilute phase of mostly assembled complexes coexists with a condensed phase of mostly free chains.}
  \label{fig:coexistence}
\end{figure}

Our aim is to compute the phase behavior of the mean-field model introduced in the previous section.
This model is completely specified by the five independent parameters $n$, $N_{\text{p}}$, $N_{\text{s}}$, $\beta\epsilon$, and $\beta\Delta G$, which are all considered to be constants when carrying out phase-coexistence calculations.
We shall assume that the number of chains per stoichiometric complex is $n=4$ throughout this work, and then consider representative scenarios for various combinations of the remaining four parameters.
Although the free-energy density, $f$, is a function of both $\phi_{\text{p}}$ and $\phi_{\text{s}}$, the concentrations of free polymers and stoichiometric complexes are related via the chemical equilibrium constraint, \eqref{eq:PSeq}.
The total polymer concentration, $\rho \equiv \rho_{\text{p}} + n \rho_{\text{s}}$, is therefore the sole independent variable when performing phase-coexistence calculations.
For convenience, we present our results in terms of a dimensionless total concentration $\rho N_{\text{p}} v_0$, which is equal to the volume fraction if all polymers are in the P state, since this dimensionless quantity is bounded by 0 and 1.
We present the composition of a solution with fixed total concentration in terms of the fraction of polymers in the S state, i.e., the ``S fraction,'' which is equal to $n \rho_{\text{s}} / \rho$.
The S fraction is also bounded by 0 and 1.

For a given set of parameters, we determine the concentrations and compositions of coexisting phases by computing and analyzing a free-energy landscape as a function of $\rho$.
This calculation is carried out in two steps.
In the first step, we determine the equilibrium S fraction and the corresponding free-energy density at a fixed value of $\rho$ by solving \eqref{eq:PSeq}.
In cases where there are multiple solutions to this chemical equilibrium equation, we choose the one with the lowest free-energy density; this choice is equivalent to performing a global minimization of $f$ as a function of the S fraction at constant $\rho$.
An example calculation showing the equivalence between constrained free-energy minimization and chemical equilibrium is illustrated in \figref{fig:coexistence}a.

In the second step, we repeat this chemical-equilibrium calculation for many values of the total concentration in order to construct the equilibrium free-energy landscape $f(\rho)$ numerically.
We then use the common-tangent construction~\cite{porter2021phase} to identify one or more coexistence regions, where two total concentrations have the same chemical potential $\mu \equiv \partial f / \partial \rho$ and osmotic pressure $P \equiv f - (\partial f / \partial \rho) \rho$, if any such regions exist.
An example landscape with two coexistence regions is illustrated in \figref{fig:coexistence}b.
In practice, we compute the convex hull of a free-energy landscape computed on a grid of discrete total concentration points, $\{(\rho, f(\rho))\}$~\cite{wolff2011thermodynamic}.
Any point $(\rho, f(\rho))$ that is not part of the convex hull must be within a coexistence region, in accordance with the common-tangent construction; adjacent grid points that are not part of the convex hull must be within the same coexistence region.
In this way, we are able to identify the number of coexistence regions and obtain estimates of the total concentrations of the coexisting phases for each region.
We then refine these estimates to compute the total concentrations and compositions of each pair of coexisting phases $\alpha$ and $\beta$.
This final refinement step is achieved by solving the equal-chemical-potential, $\mu^{(\alpha)} = \mu^{(\beta)}$, and equal-osmotic-pressure, $P^{(\alpha)} = P^{(\beta)}$, conditions to numerical precision.
The results of these calculations are most easily visualized by examining the grand-potential density, $f - \mu\rho$, with $\mu$ chosen to be equal to the chemical potential of the two coexisting phases.
The grand-potential density is equal to the osmotic pressure at equilibrium.
For example, the case shown in \figref{fig:coexistence}c corresponds to the lower-concentration coexistence region found in \figref{fig:coexistence}b, in which a dilute solution consisting primarily of assembled stoichiometric complexes coexists with a condensed phase of disassembled polymers stabilized by their associative interactions.

In order to summarize the dependence of the phase behavior on the parameters $\beta\epsilon$ and $\beta\Delta G$, we introduce the concept of a \textit{master phase diagram}.
Specifically, for fixed choices of $N_{\text{p}}$ and $N_{\text{s}}$, we determine the number of distinct coexistence regions that exist over the complete range of total concentrations, $0 \le \rho N_{\text{p}} v_0 \le 1$, as a function of $\beta\epsilon$ and $\beta\Delta G$.
Example master phase diagrams are shown in \figsref{fig:conserved-volume}a and \ref{fig:nonconserved-volume}a, and are discussed in the following sections.
In this representation, the boundaries between parameter regimes in the $\beta\epsilon$--$\beta\Delta G$ plane that have differing numbers of coexistence regions indicate lines of either first-order or second-order transitions where coexistence regions terminate.
In other words, crossing one of these boundaries on the master phase diagram means that one or more coexistence regions either appear or disappear within some range(s) of total concentration as $\beta\epsilon$ and/or $\beta\Delta G$ are tuned.

Master phase diagrams are useful for understanding how different relationships between the association and assembly free energies can result in qualitatively different phase diagrams when calculated as a function of a single control parameter.
For example, consider a biopolymer solution in which temperature controls both $\beta\epsilon$ and $\beta\Delta G$.
Depending on the precise chemical details of the biopolymers, changing temperature traces out a curve in the $\beta\epsilon$--$\beta\Delta G$ plane (see, e.g., \figsref{fig:conserved-volume}a and \ref{fig:nonconserved-volume}a).
Computing the phase behavior as a function of the total concentration at points along this temperature-parameterized curve therefore allows us to construct a conventional temperature--concentration phase diagram (see, e.g., \figsref{fig:conserved-volume}b and \ref{fig:nonconserved-volume}b).
The topology of this conventional phase diagram is determined by the way in which the temperature-parameterized curve intersects the boundaries between parameter regimes with different numbers of coexistence regions in the $\beta\epsilon$--$\beta\Delta G$ plane.
Importantly, this approach allows us to disentangle the predictions of the mean-field model, which depend on the free-energy parameters $\beta\epsilon$ and $\beta\Delta G$, from system-specific parameterizations of these quantities.

\subsection{Self-assembly with conserved excluded volume}
\label{sec:conserved-volume}

We first consider the scenario where the self-assembly of $n$ polymers into a stoichiometric complex conserves the excluded volume, such that $v_{\text{s}} = n v_{\text{p}}$.
Here we compute phase diagrams for an example case with $N_{\text{p}} = 6$ blobs per polymer chain.
Conservation of excluded volume thus implies that $N_{\text{s}} = nN_{\text{p}} = 24$, since we assume that $n=4$ chains comprise each stoichiometric complex.

\begin{figure*}[t]
\includegraphics[width=\textwidth]{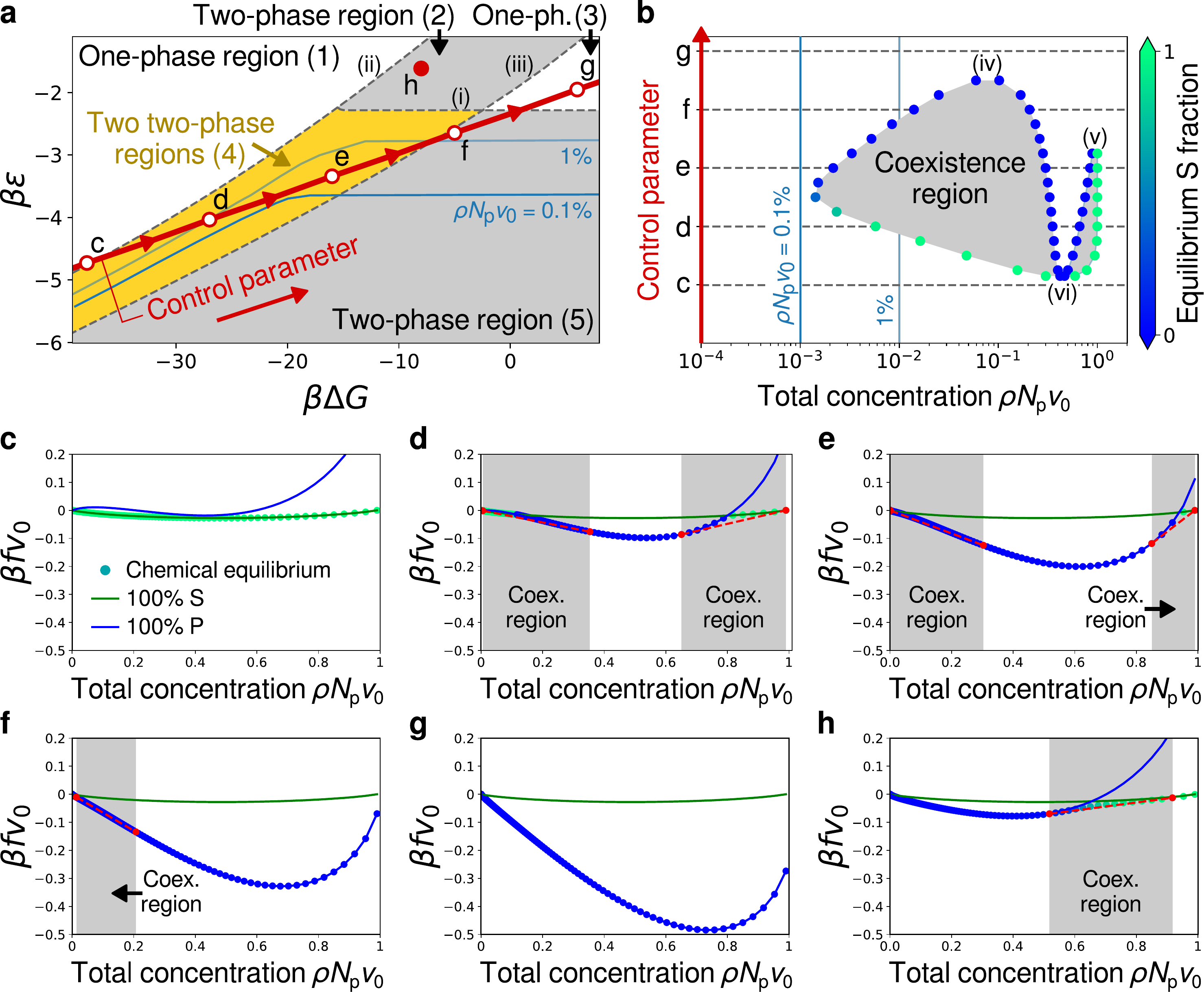}
\caption{\textbf{Phase-behavior predictions when self-assembly conserves the excluded volume of the constituent chains.}
  (a)~The master phase diagram predicted by our model exhibits rich phase behavior.  The parameter space is shaded according to the number of distinct coexistence regions at a given $\beta\epsilon$ and $\beta\Delta G$.  In the unshaded regions, the system is homogeneous at all concentrations.  Gray shaded regions indicate that phase separation occurs in a single coexistence region, while the gold region indicates the existence of two distinct coexistence regions within different concentration ranges.  These regions are bounded by dashed curves (i), (ii), and (iii), as described in the text.  Solid blue curves show where the dilute phase (of the lower-concentration coexistence region, if there are two) has the indicated total concentration.  The path taken by a hypothetical control parameter, which relates $\beta\epsilon$ to $\beta\Delta G$, is shown by a solid red line; labeled points correspond to the free-energy profiles in panels \textbf{c}--\textbf{h}.
  (b)~The phase diagram mapped out by the hypothetical control parameter shown in panel \textbf{a}.  Points representing the concentrations of coexisting phases are colored according to the equilibrium composition of stoichiometric complexes.  The coexistence regions terminate at the points (iv), (v), and (vi), as described in the text.
  (c)--(g)~Representative free-energy profiles (points) along the path taken by the control parameter shown in panel \textbf{a}.  For comparison, we also show the free-energy profiles of pure chains (100\% P, solid blue curves) and pure complexes (100\% S, solid green curves).  Red dashed lines indicate common-tangent constructions, red points indicate coexisting concentrations, and gray shaded regions indicate coexistence regions.
  (h)~A representative free-energy profile in two-phase region (2) of the master phase diagram, panel \textbf{a}.}
  \label{fig:conserved-volume}
\end{figure*}

The master phase diagram for this scenario is shown in \figref{fig:conserved-volume}a.
The model predicts five parameter regimes in the $\beta\epsilon$--$\beta\Delta G$ plane: two distinct regimes, (1) and (3), in which only one phase exists at all concentrations; two distinct regimes, (2) and (5), in which a single coexistence region exists over a finite range of concentrations; and one regime, (4), in which two different coexistence regions exist over different concentration ranges.
We discuss the physical origin and significance of these various regimes in the remainder of this section.

\subsubsection{Phase diagram construction as a function of a control parameter}

As discussed in \secsref{sec:model} and \ref{sec:coexistence}, the free-energy parameters $\beta\epsilon$ and $\beta\Delta G$ must be related to one another due to their common dependence on the interactions between pairs of binding sites (\figref{fig:model}a).
We therefore construct a conventional concentration-dependent phase diagram by tracing a path through the $\beta\epsilon$--$\beta\Delta G$ plane that depends on a control parameter.
For example, if the control parameter is temperature, then the resulting conventional phase diagram depicts the phase behavior in the temperature--concentration plane.
We show a possible parameterization of $\beta\epsilon$ and $\beta\Delta G$ in \figref{fig:conserved-volume}a, which results in the conventional phase diagram shown in \figref{fig:conserved-volume}b.
This example curve is chosen because it passes through four of the five regimes on the master phase diagram.
Importantly, this example also satisfies the relation $\partial \beta\epsilon / \partial \beta\Delta G > 0$, since both free energies should typically be expected to increase or decrease in tandem.
Specific points along the parameterized curve are labeled c--g in \figref{fig:conserved-volume}a and are indicated on the conventional phase diagram in \figref{fig:conserved-volume}b.

The appearance and disappearance of various coexistence regions in \figref{fig:conserved-volume}b is determined by the intersections between the control-parameter curve and the boundaries of the distinct regimes in \figref{fig:conserved-volume}a.
Starting from low values of the control parameter, the polymer solution is initially in a one-phase region at all concentrations at point (c).
Upon crossing line (ii) in \figref{fig:conserved-volume}a, two distinct coexistence regions appear simultaneously at point (vi) in \figref{fig:conserved-volume}b.
These coexistence regions imply that a polymer solution at a total concentration within a shaded region will phase separate at equilibrium into two coexisting phases with different concentrations and compositions, as indicated by the equilibrium S fraction shown on the coexistence-region boundaries in \figref{fig:conserved-volume}b.
Thus, while the number of coexistence regions is the same at points (d) and (e), phase coexistence in the lower-concentration coexistence region is qualitatively different at these two points.
Specifically, the dilute phase at point (d) is dominated by assembled stoichiometric complexes, while the dilute phase at point (e) is primarily composed of free polymers.
The control-parameter curve then crosses line (iii) in \figref{fig:conserved-volume}a, which indicates the termination of the high-concentration coexistence region at point (v) in \figref{fig:conserved-volume}b, leaving only the lower-concentration coexistence region at point (f).
Finally, the control-parameter curve crosses line (i) in \figref{fig:conserved-volume}a, corresponding to point (iv) in \figref{fig:conserved-volume}b, so that the polymer solution exists in a single phase at all concentrations once again at point (g).

Both the higher-concentration and the lower-concentration coexistence regions in \figref{fig:conserved-volume}b exhibit re-entrance, since a polymer solution at constant total concentration can pass from a single phase to two phases and back to a single phase as the control parameter is varied monotonically from point (c) to point (g).
However, re-entrance only occurs over a finite range of total concentrations.
To determine whether re-entrance will occur at a specific total concentration, we also plot curves of constant dilute-phase concentration on the master phase diagram in \figref{fig:conserved-volume}a.
These curves indicate the conditions in the $\beta\epsilon$--$\beta\Delta G$ plane at which the dilute phase has the indicated total concentration, and thus where the binodal intersects a line of fixed total concentration in a control parameter versus total concentration phase diagram (e.g., \figref{fig:conserved-volume}b).
Because the example control-parameter curve shown in \figref{fig:conserved-volume}a crosses the curve of a 1\% dilute-phase concentration twice but does not cross the curve of a 0.1\% dilute-phase concentration, we can easily predict that re-entrance will occur for $\rho N_{\text{p}} v_0 = 1\%$ but not at $\rho N_{\text{p}} v_0 = 0.1\%$ in \figref{fig:conserved-volume}b.
These examples demonstrate how the essential features of the conventional phase diagram can be inferred graphically by comparing a control-parameter curve, which is specific to a particular system, with the master phase diagram, which depends only on the parameters $N_{\text{p}}$, $N_{\text{s}}$, and $n$ in this model.

\subsubsection{Free-energy landscapes}

Representative free-energy landscapes at various points in the $\beta\epsilon$--$\beta\Delta G$ plane, corresponding to each labeled point on the control-parameter curve as well as the additional labeled point (h), are shown in \figref{fig:conserved-volume}c--h.
Equilibrium landscapes are indicated by points whose coloring reflects the composition, as in \figref{fig:conserved-volume}b.
Coexistence regions and tie-line constructions are indicated for each equilibrium landscape.
For comparison, we also show free-energy landscapes computed assuming a \textit{nonequilibrium} composition of either 100\% S or 100\% P.

We find that we can predict the qualitative phase behavior at each point along the control-parameter curve in \figref{fig:conserved-volume}b by examining the relationship between these nonequilibrium landscapes, despite the fact that the composition is never exactly 100\% S or 100\% P due to the entropy of mixing.
In \figref{fig:conserved-volume}c, the equilibrium landscape is close to the 100\% S curve, which is always below the 100\% P curve.
Thus, the polymer solution primarily consists of assembled stoichiometric complexes in a single phase at all concentrations.
In \figref{fig:conserved-volume}d, the lower-concentration coexistence region is approximately given by a common-tangent construction between the 100\% S curve (on the left) and the 100\% P curve (on the right).
However, the lower-concentration coexistence region in \figref{fig:conserved-volume}e and the sole coexistence region in \figref{fig:conserved-volume}f are approximately given by a common-tangent construction using the 100\% P curve alone, since this curve is nonconvex due to the associative contribution to the free energy, \eqref{eq:fassoc}.
This nonconvexity in the 100\% P curve disappears at high $\beta\epsilon$, resulting in no phase separation in \figref{fig:conserved-volume}g.
By contrast, the higher-concentration coexistence region in \figsref{fig:conserved-volume}d and \ref{fig:conserved-volume}e and the sole coexistence region in \figref{fig:conserved-volume}h are approximately given by a common-tangent construction between the 100\% P curve (on the left) and the 100\% S curve (on the right).

\subsubsection{Parameter-regime boundaries on the master phase diagram}

The nature of the boundaries between regimes in the master phase diagram can be understood by examining the behavior of the system at points where the coexistence regions terminate in \figref{fig:conserved-volume}b.
Point (iv) in \figref{fig:conserved-volume}b is a critical point, since the derivatives $\partial\mu/\partial\rho$ and $\partial^2\mu/\partial\rho^2$ vanish.
As noted at the end of \secref{sec:model}, this critical point is typically referred to as an upper critical solution temperature (UCST).
Line (i) in \figref{fig:conserved-volume}a is therefore a line of critical points that bound the lower-concentration coexistence region.
Since the two phases that merge at point (iv) are composed mostly of free polymers, line (i) is nearly independent of $\beta\Delta G$.
The value of $\beta\epsilon$ along this line is thus approximately equal to the critical association free energy of an associating polymer solution in the absence of self-assembly (i.e., $\beta\Delta G \rightarrow \infty$).

By contrast, point (v) in \figref{fig:conserved-volume}b is a first-order transition, where the composition changes discontinuously from mostly assembled stoichiometric complexes to mostly free polymers upon a change in $\beta\epsilon$ and/or $\beta\Delta G$.
This transition occurs when the free energy of a 100\% P solution drops below that of a 100\% S solution at $\rho N_{\text{p}} v_0 = 1$, which we can see by comparing the free-energy landscapes shown in \figsref{fig:conserved-volume}e and \ref{fig:conserved-volume}f.
We can therefore predict the line of first-order transitions indicated by line (iii) in \figref{fig:conserved-volume}a by equating the free energies of 100\% P and 100\% S solutions at $\rho N_{\text{p}} v_0 = 1$.
To the left of line (iii), both coexisting phases have total concentrations less than one, as evidenced by the example coexistence region shown in \figref{fig:conserved-volume}h.
To the right of line (iii), the higher-concentration coexistence region does not exist.

\subsubsection{Physical origin of re-entrance}

We now focus on the physical origin of the re-entrant behavior that occurs at relatively low values of the total concentration in \figref{fig:conserved-volume}b.
In essence, re-entrance manifests because a solution of free polymers phase separates at high values of the control parameter, but a single-phase solution of predominantly assembled stoichiometric complexes becomes stable at low values of the control parameter.
Re-entrance therefore results from mutually exclusive competition between polymer phase separation and stoichiometric complex self-assembly.
More precisely, re-entrance occurs when the control-parameter curve crosses a line of constant dilute-phase concentration (e.g., $\rho N_{\text{p}}v_0 = 1\%$) \textit{twice} in \figref{fig:conserved-volume}a.
These constant dilute-phase concentration curves are bounded by lines (i) and (ii), so this form of re-entrance can only appear in a limited portion of the $\beta\epsilon$--$\beta\Delta G$ plane.
While the upper bound of the lower-concentration coexistence region with respect to the control parameter is set by point (iv) in \figref{fig:conserved-volume}b, the lower bound on this lower-concentration coexistence region is determined by point (vi).
This lower bound corresponds to the intersection between line (ii) and the control-parameter curve in \figref{fig:conserved-volume}a.

\begin{figure}[htb!]
\includegraphics[width=\columnwidth]{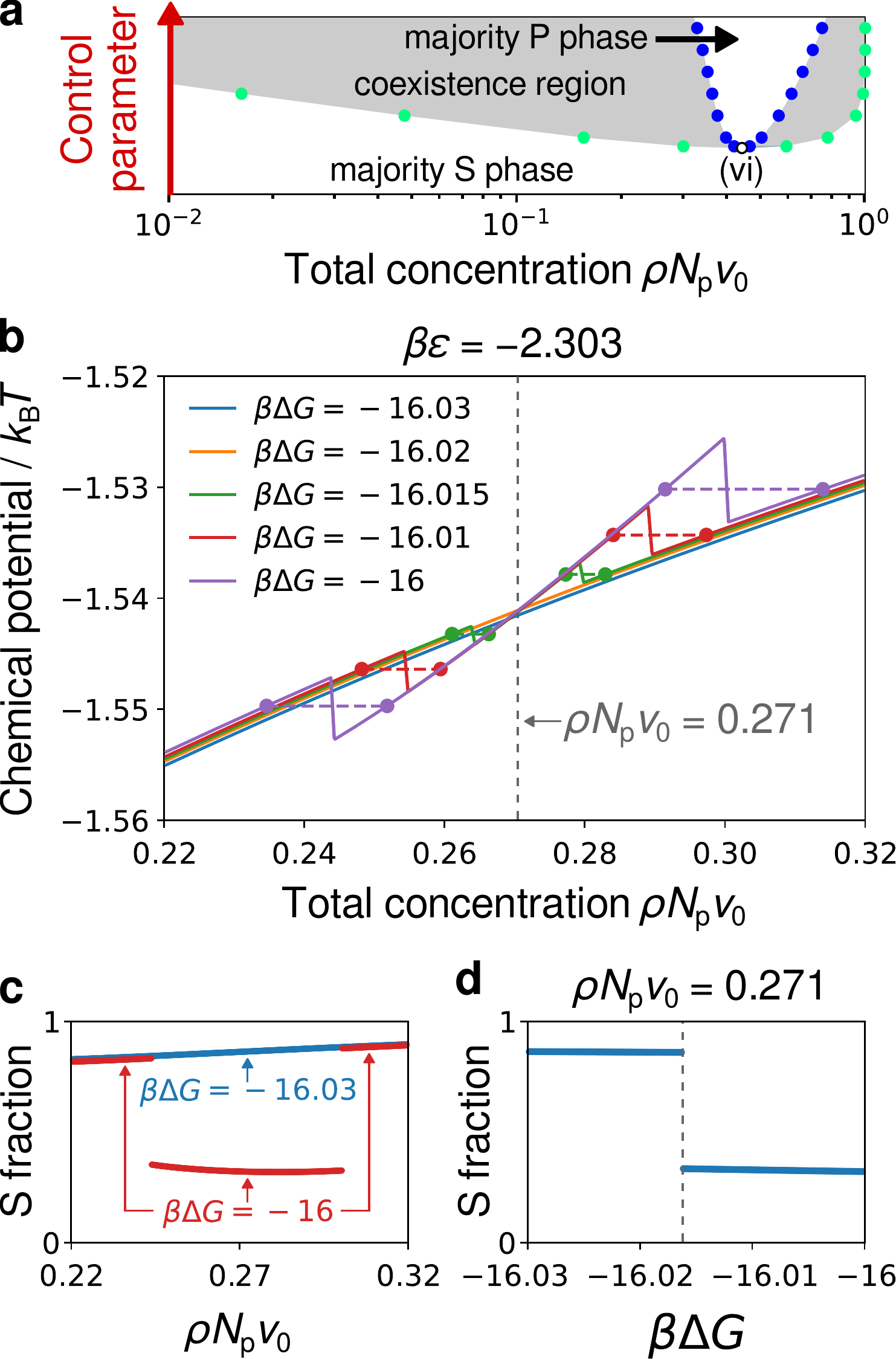}
\caption{\textbf{Coexistence regions terminate at a first-order phase transition in the conserved-excluded-volume scenario.}
  (a)~Here we examine the nature of the transition from a one-phase region to two two-phase regions along line (ii) in \figref{fig:conserved-volume}a, which corresponds to point (vi) in this zoomed-in portion of the phase-diagram shown in \figref{fig:conserved-volume}b.
  (b)~The concentration-dependent chemical potential, $\mu$, at fixed $\beta\epsilon=-2.303$ and variable $\beta\Delta G$.  Coexisting phases are indicated by circles and connected by dashed lines.  The transition from two coexistence regions ($\beta\Delta G \gtrsim -16.015$) to zero coexistence regions ($\beta\Delta G \lesssim -16.02$) does not involve a critical point, which would imply a concentration where $\partial \mu / \partial \rho = 0$.  Instead, the two coexistence regions move towards one another as $\beta\Delta G$ is decreased and vanish at the azeotrope concentration $\rho N_{\text{p}} v_0 \simeq 0.271$ (dashed line), where $\partial \mu / \partial \rho$ jumps discontinuously.
  (c)~When no coexistence regions are present ($\beta\Delta G = -16.03$), the composition is a continuous function of $\rho$.  Yet when two coexistence regions are present ($\beta\Delta G = -16$), the composition changes discontinuously across each coexistence region.
  (d)~The first-order nature of this transition at finite $\beta\Delta G$ (dashed line), corresponding to line (ii) in \figref{fig:conserved-volume}, is clear from the discontinuity in the composition at the azeotrope concentration.  Moving from left to right across this plot is analogous to moving through point (vi) in \figref{fig:conserved-volume}b at constant $\rho$ from below to above.}
  \label{fig:first-order}
\end{figure}

Yet unlike point (iv) in \figref{fig:conserved-volume}b, point (vi) is \textit{not} a critical point.
Instead, point (vi) indicates a first-order transition where the composition jumps from a high value to a low value of the equilibrium S fraction as the control parameter is increased.
The distinction between the behavior at point (vi) and a lower critical solution temperature (LCST) can be understood by examining the chemical potential in the vicinity of the transition (\figref{fig:first-order}a).
In \figref{fig:first-order}b, we show the chemical potential as a function of the total concentration at fixed $\beta\epsilon$.
As we tune $\beta\Delta G$ near the transition point, we see that the two coexistence regions converge at a crossover concentration (vertical dashed line in \figref{fig:first-order}b).
However, the derivative of the chemical potential at the crossover concentration does not vanish, as would be required for a second-order transition.
By contrast, the equilibrium S fraction at the crossover concentration changes discontinuously, as shown in \figref{fig:first-order}c--d.
Since the composition is determined from a first derivative of the free energy, this behavior is consistent with a first-order transition.

We propose that this behavior can be understood intuitively by drawing an analogy to a positive azeotrope~\cite{shephard2016microstructures}.
In a binary mixture of two fluids, a positive azeotrope occurs at a specific composition at which the two fluids share the same boiling point; thus, raising the temperature of the mixture at the azeotrope composition causes the liquid mixture to transform directly into a mixture of gases with the same composition.
Qualitatively, the free-energy landscape of a fluid mixture with an azeotrope is analogous to that of the nonequilibrium 100\% S and 100\% P curves shown in \figref{fig:conserved-volume}c--d, whereby the ``liquid-like'' 100\% S curve is stable below a transition point $(\beta\Delta G,\beta\epsilon)$ and the ``vapor-like'' 100\% P curve is stable above.
We therefore propose that the crossover concentration shown in \figref{fig:first-order}b is analogous to an azeotrope composition, even though there is technically only one equilibrium free-energy landscape as a function of the total concentration in our model due to the chemical equilibrium constraint, \eqref{eq:PSeq}.
Finally, we note that line (i) terminates at line (ii) in \figref{fig:conserved-volume}a for a similar reason: For sufficiently negative $\beta\Delta G$, the 100\% S free-energy landscape curve becomes more stable than the 100\% P curve at all concentrations.
Consequently, the nonconvex portion of the free-energy landscape, which describes a polymer solution composed of mostly free polymers, becomes metastable to the left of line (ii) in \figref{fig:conserved-volume}a.

\subsubsection{Dependence of the master phase diagram on $N_{\text{p}}$ and $n$}

Up to this point, we have discussed the master phase diagram assuming the specific parameter choices $N_p=6$ and $n=4$.
These calculations can easily be repeated using alternative parameters.
In this way, we find that the topology of the master phase diagram remains unchanged as long as $n > 1$ and the excluded volume is conserved, such that $N_{\text{s}} = nN_{\text{p}}$.
However, changing the parameters $N_{\text{p}}$ and $n$ alter the positions of the phase boundaries in the $\beta\epsilon$--$\beta\Delta G$ plane, as we discuss next.

The line of critical points (i) is nearly independent of the degree of complexation $n$ because these critical points only depend on the properties of free polymers, except when in close proximity to the intersection of lines (i) and (ii).
Line (i) shifts upward in the $\beta\epsilon$--$\beta\Delta G$ plane as the degree of polymerization $N_{\text{p}}$ increases, since the critical per-binding-site association free energy becomes weaker as the number of binding sites per polymer increases.
This behavior is analogous to that of the Flory--Huggins homopolymer model, in which the critical interaction strength weakens as the degree of polymerization increases~\cite{rubinstein2003polymer}.

By contrast, the two first-order phase boundaries, lines (ii) and (iii), depend on both $n$ and $N_{\text{p}}$.
The slope of line (iii) is exactly proportional to $nN_{\text{p}}$; this can be shown by equating the free energies of the 100\% P and 100\% S solutions at $\rho N_{\text{p}} v_0 = 1$.
The slope of line (ii), as well as the difference in $\beta\Delta G$ between lines (ii) and (iii) at constant $\beta\epsilon$, also scales roughly linearly with $nN_{\text{p}}$.
This dependence can be understood by considering the transition between regions (1) and (4) in \figref{fig:conserved-volume}a in the strong association limit, $\beta\epsilon \rightarrow -\infty$~\cite{hegde2023competition}.
In this limit, the relative stabilities of the 100\% P and 100\% S solutions are primarily determined by the balance between the association free energy in the P state, in which nearly all binding sites are associated since $X_{\text{p}} \rightarrow 0$, and the internal contribution to the free energy in the S state, which is nearly proportional to $\beta\Delta G$.
Consequently, $\beta\Delta G$ must scale with the total number of associative interactions in which the complex-forming polymers could potentially engage, which is equal to $nN_{\text{p}}$.
Thus, $nN_{\text{p}}$ sets the scale for both lines (ii) and (iii).
Increasing $nN_{\text{p}}$ tends to stretch the master phase diagram in the horizontal direction.

\subsection{Self-assembly with nonconserved excluded volume}
\label{sec:nonconserved-volume}

Next, we consider the alternative case in which self-assembly does not conserve the excluded volume.
This scenario can be motivated by considering the second-virial coefficient between two assembled stoichiometric complexes in solution, which is given by $B_{\text{ss}} = N_{\text{s}}v_0/2$ in our model (see \appref{app:B2}).
This second-virial coefficient represents the total volume (divided by two) that the center of mass of one stoichiometric complex cannot access due to the presence of the other complex.
If the polymers self-assemble into a structure that has a lower density than a polymer melt, then it is possible to have $v_{\text{s}} = 2B_{\text{ss}} > n v_{\text{p}}$.
For example, this scenario is relevant in the case of nucleic-acid complexes, because secondary-structure formation increases the persistence length of duplexes relative to free strands~\cite{rovigatti2014accurate,conrad2022emulsion}; nucleic-acid complexes can therefore have a relatively low density of polymer within a relatively large pervaded volume.
We therefore consider an example scenario in which $N_{\text{s}} = 200 > nN_{\text{p}} = 24$ in this section.

\begin{figure*}[t]
\includegraphics[width=\textwidth]{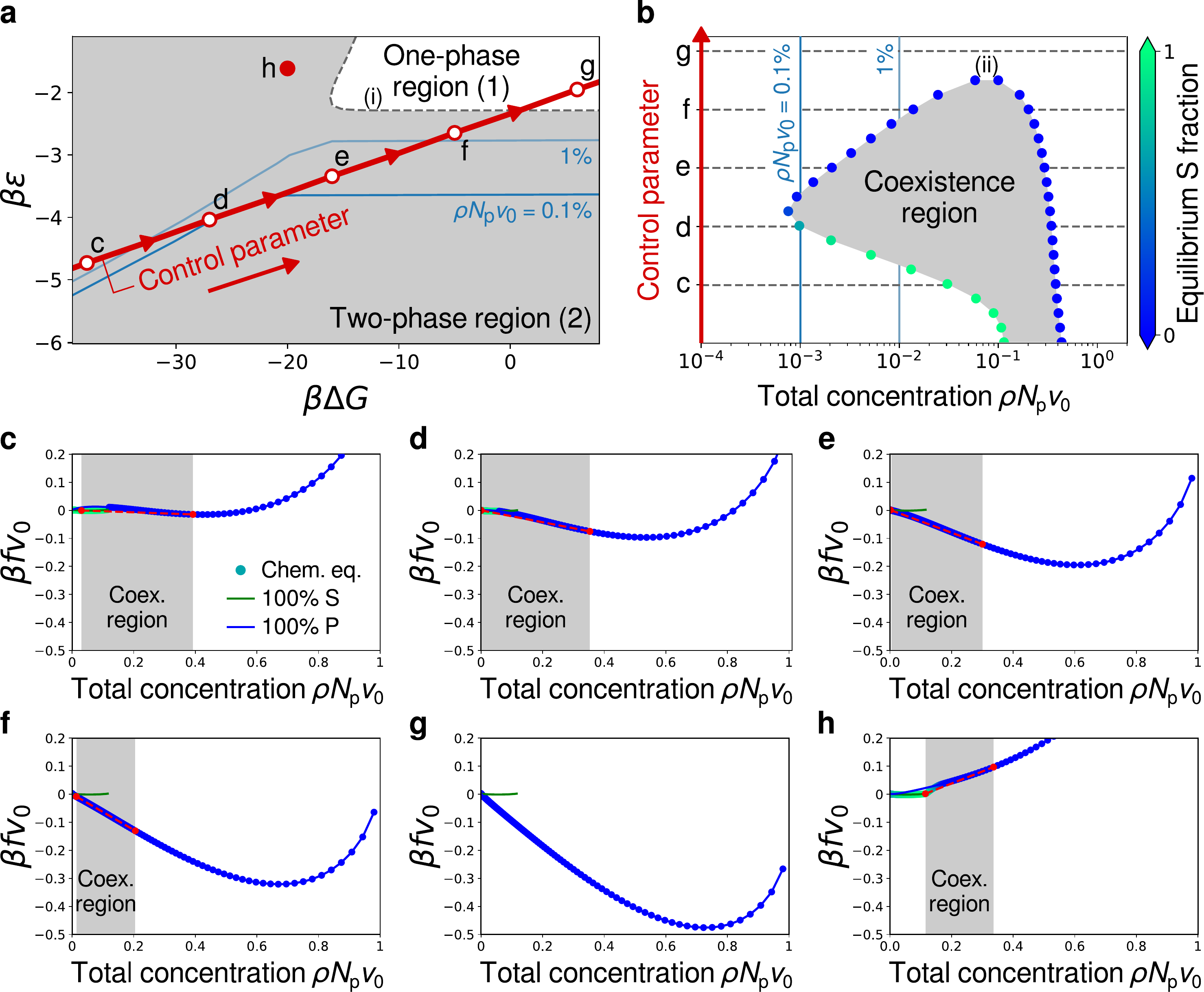}
\caption{\textbf{Phase-behavior predictions when self-assembly does not conserve the excluded volume of the constituent chains.}
  (a)~A single one-phase region (unshaded), where the system is homogeneous at all concentrations, lies to the right of dashed curve (i) and above dashed curve (ii), as described in the text.  A two-phase region exists within some concentration range everywhere else within the master phase diagram.  See \figref{fig:conserved-volume}a for explanations of the control parameter path and total concentration curves.
  (b)~When the excluded volume is not conserved by the two-state model of self-assembly, the lone coexistence region terminates at a single point (iii), as described in the text.  This phase diagram is constructed analogously to that shown in \figref{fig:conserved-volume}b.
  (c)--(h): Representative free-energy profiles along the path taken by the control parameter shown in panel \textbf{a}, and at the additional labeled point (h) in panel \textbf{a}.  See the description of \figref{fig:conserved-volume}c--h for details.  Because of the excluded volume difference, the 100\% S curves terminate at the total concentration where the complexes are close-packed, $\rho N_{\text{s}} v_0 / n = 1$, where $n=4$ in this example.  Consequently, condensed phases with $\rho N_{\text{p}} v_0 > n N_{\text{p}} / N_{\text{s}}$ are always polymer-dominated.}
  \label{fig:nonconserved-volume}
\end{figure*}

\subsubsection{Similarities and differences with the conserved-excluded-volume scenario}

The master phase diagram for this scenario is shown in \figref{fig:nonconserved-volume}a.
The control-parameter curve, which is the same as that in \figref{fig:conserved-volume}a, maps out the conventional concentration-dependent phase diagram shown in \figref{fig:nonconserved-volume}b.
In contrast with the conserved-excluded-volume scenario, this master phase diagram only exhibits two parameter regimes: one regime, (1), in which the solution exists in a single phase at all concentrations; and one regime, (2), with a single coexistence region.
Yet despite these drastic changes in the topology of the master phase diagram relative to the conserved-excluded-volume scenario, the coexistence region in \figref{fig:nonconserved-volume}b strongly resembles the lower-concentration coexistence regime in \figref{fig:conserved-volume}b for values of the control parameter above point~(c).

These differences can be understood by noting that the increased excluded volume of the stoichiometric complex means that chains must exist in the free-polymer state at high total concentrations.
More precisely, since the volume fraction $\phi_{\text{s}}$ cannot exceed one, the maximum total concentration at which stoichiometric complexes can exist is given by $\rho N_{\text{p}} v_0 = n N_{\text{p}} / N_{\text{s}}$.
This upper limit on the concentration of stoichiometric complexes is reflected in the free-energy landscapes shown in \figref{fig:nonconserved-volume}c--h, where the 100\% S landscapes do not extend over the full range of total concentrations.
Consequently, there is no analog of a higher-concentration coexistence region, as observed in the conserved-excluded-volume case, when we consider the nonconserved scenario.
However, the phase behavior at low total concentrations, $\rho N_{\text{p}} v_0 \ll n N_{\text{p}} / N_{\text{s}}$, is qualitatively the same in both excluded-volume scenarios.
We also note that the boundary between the two regimes, line (i) in \figref{fig:nonconserved-volume}a, is a single, continuous line of critical points.
This critical line is the same as line (i) in the conserved-excluded-volume case at large $\beta\Delta G$, but turns into a continuous transition between assembled stoichiometric complexes and free polymers when it bends upward near point (h) in \figref{fig:nonconserved-volume}a.

\subsubsection{Physical origin of re-entrance}

Re-entrance occurs in the nonconserved-excluded-volume scenario due to the same competition between self-assembly and polymer phase separation discussed in the context of the conserved-excluded-volume scenario.
Minor quantitative differences, such as the maximum total concentration at which re-entrance occurs, appear between the two scenarios due to the differences in the excluded-volume and internal contributions to the free energy, $f^{\text{ev}}$ and $f^{\text{int}}$, in \eqsref{eq:fev} and \ref{eq:fint}.
However, the absence of a one-phase parameter regime at low $\beta\Delta G$ in the nonconserved case means that the coexistence region does not terminate at low values of the control parameter in \figref{fig:nonconserved-volume}b.
Instead, the maximum concentration at which re-entrance can occur tends to $\phi_{\text{s}} = 1$ in the limit $\beta\Delta G \rightarrow -\infty$.
In this limit, phase coexistence occurs between a lower-concentration phase composed entirely of assembled stoichiometric complexes and a higher-concentration phase of free polymers, regardless of the value of $\beta\epsilon$.

\section{Discussion}
\label{sec:discussion}

We have demonstrated how competition between self-assembly and liquid--liquid phase separation can emerge from the same set of molecular interactions among biopolymers in solution.
Our theoretical approach captures the many-body interactions that are inherent to this competition by coarse-graining the conformational states of polymers into either free chains and assembled stoichiometric complexes.
The resulting mean-field model predicts that re-entrance can emerge from the competition between self-assembly and phase separation, both of which involve only attractive associative interactions.
Our results provide insight into the key features that control the phase behavior of these systems, providing a unified view of different phase diagrams that can result given a parametric dependence on temperature, ionic strength, pH, or other experimentally controllable variables.

Importantly, we find that re-entrance in this model is not a consequence of a lower critical solution temperature (LCST), despite qualitative similarities between the phase diagrams predicted by our model and those involving UCST+LCST miscibility loops~\cite{rubinstein2003polymer,rovigatti2013self}.
In the conserved-excluded-volume scenario, the lower termination of the coexistence regions occurs at a first-order transition, which is qualitatively different from an LCST.
By contrast, when self-assembly does not conserve excluded volume, the coexistence region does not terminate at all in the limit of stable stoichiometric complexes.
Nonetheless, the physical origin of re-entrance is the same in both cases, and the phase behavior at low polymer concentrations is essentially unchanged between these two scenarios.
We therefore believe that the qualitative predictions of our model---particularly the absence of an LCST---are robust despite the approximations invoked in our mean-field treatment.
We note that analogous behavior has also been observed in conceptually similar mean-field models of monomeric as opposed to polymeric solutions, in which phase separation either competes with self-assembly~\cite{reinhardt2011re} or a unimolecular chemical reaction~\cite{bartolucci2021controlling}.
In that case of unimolecular chemical reactions~\cite{bartolucci2021controlling}, however, the low-temperature terminus of the coexistence region [i.e., point (vi) in \figref{fig:conserved-volume}b] always occurs at $\phi = 1$.

A key approximation in this work is the use of a two-state model of self-assembly.
In principle, the accuracy of our model could be improved by accounting for additional intermediate states between completely dissociated polymers and completely assembled stoichiometric complexes.
Our model also assumes that the excluded volume of the stoichiometric complexes, $v_{\text{s}}$, is a constant.
A more versatile model could account for variations in $v_{\text{s}}$ as a function of concentration, which would provide a better description of complexes that deform at high concentrations in the nonconserved-excluded-volume scenario.
However, we do not believe that these modifications would affect our qualitative conclusions regarding the nature of the various phase transitions.
In particular, the first-order transitions that we observe emerge, fundamentally, from the cooperative nature of the self-assembly process, which implies the existence of a free-energy barrier between the free-polymer (P) and assembled stoichiometric-complex (S) states.
We therefore believe that the phase behavior will remain qualitatively unchanged as long as this essential feature is preserved.
For the same reason, we believe that this model can apply to a wide variety of self-assembling polymer solutions in which self-assembly is cooperative and mutually exclusive with phase separation.
Our approach would be less appropriate for modeling the phase behavior of polymer solutions in which less cooperative intramolecular bonding competes with associative polymer phase separation, as is the case in many experiments on nucleic-acid repeat sequences~\cite{kimchi2023uncovering}.

In conclusion, we have introduced and analyzed a broadly applicable mean-field model of biopolymer assembly and phase separation.
Our ``master-phase-diagram'' approach makes it possible to derive system-specific phase diagrams in a unified and intuitive fashion.
It is also straightforward to extend our theoretical framework to model more complex scenarios involving ``hierarchical'' assembly and phase separation, such as polymer solutions in which self-assembled complexes can themselves undergo phase separation via associative interactions~\cite{hegde2023competition}.
Most importantly, the simple manner in which our model can be parameterized makes this framework useful for generating experimentally testable predictions for complex biopolymer solutions.

\begin{acknowledgments}
  This work was supported by a grant from the National Science Foundation (DMR-2143670) to WMJ, as well as support from the Human Frontier Science Program (RGP0029) and the Smith Family Foundation to WBR.
  Source code is provided at \texttt{https://github.com/wmjac/}\\\texttt{polymer-phase-sep-self-assembly}.
\end{acknowledgments}


\appendix
\setcounter{figure}{0}
\renewcommand{\thefigure}{A\arabic{figure}}

\section{Excluded volume formulae}
\label{app:EFH}

\begin{figure}
  \includegraphics[width=8.5cm]{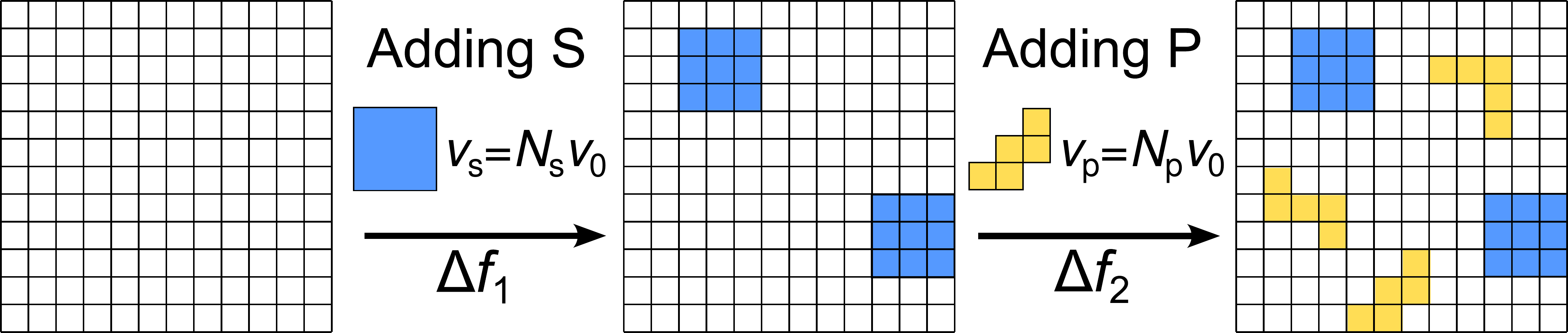}
  \caption{Two-step calculation of the excluded-volume contribution to the free energy.  Starting from the reference state of an empty lattice (state 0, left), stoichiometric complexes (S) are added in step one.  Stoichiometric complexes occupy $N_{\text{s}}$ \textit{contiguous} lattice sites.  The free-energy change associated with this process is $\Delta f_1$, resulting in the intermediate configuration (state 1, middle).  In step two, polymers (P) are added to reach the final configuration (state 2, right).  Polymers occupy $N_{\text{p}}$ \textit{non-contiguous} lattice sites.  The free-energy change associated with this second process is $\Delta f_2$.}
  \label{fig:EFH}
 \end{figure}

To derive an expression for the excluded-volume contribution to the mixing free energy, we consider a cubic lattice with $N$ lattice sites.
Each lattice site has the same volume, $v_0$, as a monomer (i.e., a blob) of the polymer, P.
Each polymer chain occupies $N_{\text{p}}$ lattice sites.
Each stoichiometric complex, referred to as an ``S molecule'' in this section, occupies an excluded volume of $v_{\text{s}} = N_{\text{s}} v_0$.
We compute the free energy of mixing, in the absence of associative interactions or chemical reactions, in a two-step manner (\figref{fig:EFH}).
An empty lattice, with a free energy equal to zero, is chosen as the reference state.
At each step, the free-energy change is purely entropic due to the increase in the number of microstates, $\Omega_{i-1} \rightarrow \Omega_i$,
\begin{equation}
\beta N \Delta f_i v_0 = - \frac{\Delta S_i}{k_B} = -\ln\left(\frac{\Omega_i}{\Omega_{i - 1}}\right),
\end{equation}
where $\Delta S_i$ is the entropy change associated with step $i$ from state $i-1$ to state $i$.

First, we consider the free-energy change due to the addition of S molecules occupying a volume fraction $\phi_{\text{s}}$.
Importantly, each S molecule is considered to be a compact, rigid unit, and must therefore occupy $N_{\text{s}}$ \textit{contiguous} lattice sites.
We therefore compute the approximate entropy change associated with step 1 (\figref{fig:EFH}) by coarse-graining the lattice into ``super sites'' of volume $v_0 N_{\text{s}}$, and then applying the standard regular-solution formula~\cite{porter2021phase},
\begin{equation}
\beta \Delta f_1 v_0 = \frac{\phi_{\text{s}}}{N_{\text{s}}} \ln \phi_{\text{s}} + \frac{1 - \phi_{\text{s}}}{N_{\text{s}}} \ln (1 - \phi_{\text{s}}).
\end{equation}
Second, we add P molecules occupying a volume fraction $\phi_{\text{p}}$ into the system.
Following the Flory--Huggins derivation~\cite{rubinstein2003polymer}, we ignore chain connectivity when assigning polymer blobs to \textit{non-contiguous} unoccupied lattice sites.
This leads to the mean-field result
\begin{equation}
\beta \Delta f_2 v_0 = \frac{\phi_{\text{p}}}{N_{\text{p}}} \ln \left( \frac{\phi_{\text{p}}}{1 - \phi_{\text{s}}} \right) + \phi_0 \ln \left( \frac{\phi_0}{1 - \phi_{\text{s}}} \right),
\end{equation}
where ${\phi_0 \equiv 1 - \phi_{\text{s}} - \phi_{\text{p}}}$ represents the volume fraction not occupied by either free polymers or stoichiometric complexes.
Summing the free-energy changes of the two steps, we find that the total free-energy difference is
\begin{align}
\beta &\left(f^{\text{id}} + f^{\text{ev}}\right) v_0 = \frac{\phi_{\text{s}}}{N_{\text{s}}} \ln \phi_{\text{s}} + \frac{1 - \phi_{\text{s}}}{N_{\text{s}}} \ln (1 - \phi_{\text{s}}) \nonumber \\
&\qquad\qquad\; + \frac{\phi_{\text{p}}}{N_{\text{p}}} \ln \left(\frac{\phi_{\text{p}}}{1 - \phi_{\text{s}}} \right) + \phi_0 \ln \left( \frac{\phi_0}{1 - \phi_{\text{s}}} \right),
\label{eq:f_EFH}
\end{align}
which accounts for both the ideal-gas and the excluded-volume contributions to the free energy.
Taking partial derivatives of  \eqref{eq:f_EFH} with respect to $\rho_{\text{p}}$ and $\rho_{\text{s}}$, and subtracting the ideal chemical potentials $\beta^{-1}\ln\phi_{\text{p}}$ and $\beta^{-1}\ln\phi_{\text{s}}$, gives the excluded-volume contributions to the excess chemical potentials,
\begin{align}
  \label{eq:mu_EFH_s}
  \beta \mu^{\text{ev}}_{\text{s}} &= -\ln (1 - \phi_{\text{s}}) + N_{\text{s}} \ln \left( \frac{1 - \phi_{\text{s}}}{\phi_0} \right) \\
  &\quad - N_{\text{s}} \left( 1 - \frac{1}{N_{\text{p}}} \right) \frac{\phi_{\text{p}}}{1 - \phi_{\text{s}}}, \nonumber \\
  \label{eq:mu_EFH_p}
  \beta \mu^{\text{ev}}_{\text{p}} &= 1 - N_{\text{p}} + (N_{\text{p}} \!-\! 1) \ln (1 \!-\! \phi_{\text{s}}) - N_{\text{p}} \ln \phi_0.
\end{align}
Both $\beta\mu^{\text{ev}}_{\text{s}}$ and $\beta\mu^{\text{ev}}_{\text{p}}$ diverge as $\phi_0 \rightarrow 0$.
Since $\phi_{\text{s}} + \phi_{\text{p}} \le 1$, this means that both $\beta\mu^{\text{ev}}_{\text{s}}$ and $\beta\mu^{\text{ev}}_{\text{p}}$ diverge if $\phi_{\text{s}} \rightarrow 1$ or $\phi_{\text{p}} \rightarrow 1$.

\section{Comparison with the Flory--Huggins polymer model}
\label{app:B2}

The excluded-volume contributions embodied in \eqsref{eq:f_EFH}, (\ref{eq:mu_EFH_s}), and (\ref{eq:mu_EFH_p}) differ from those implied by the standard Flory--Huggins free-energy density~\cite{rubinstein2003polymer},
\begin{equation}
  \label{eq:f_FH}
  \beta f^{\text{FH}} v_0 = \frac{\phi_{\text{s}}}{N_{\text{s}}} \ln \phi_{\text{s}} + \frac{\phi_{\text{p}}}{N_{\text{p}}} \ln \phi_{\text{p}} + \phi_0 \ln \phi_0,
\end{equation}
which treats both the S and P species as mean-field polymers.
The excluded-volume contribution to the chemical potential of species $i \in \{\text{s},\text{p}\}$ is
\begin{equation}
  \label{eq:mu_FH}
  \beta \mu^{\text{FH,ev}}_i = 1 - N_i - N_i \ln \phi_0
\end{equation}
in the Flory--Huggins model.
Focusing on the case of the S molecule, \eqref{eq:mu_FH} highlights the problem with applying the Flory--Huggins model to our system: The chemical potential of the S molecule depends on $N_{\text{s}}$ regardless of the polymer concentration.
As a result, the Flory--Huggins equation of state of a pure solution of S molecules depends on the ratio of the excluded volume of a rigid S molecule to the excluded volume of a polymer blob, even when no polymer is present in the system.
This is an undesirable feature, which necessitates the use of the formulae derived in \appref{app:EFH}.

It is also instructive to compare the second-virial coefficients, $B_{\text{ss}}$, $B_{\text{sp}}$, and $B_{\text{pp}}$, of the two models.
To this end, we expand the pressure, $P = -f + \mu_{\text{s}} \rho_{\text{s}} + \mu_{\text{p}} \rho_{\text{p}}$, to second order in the concentrations $\rho_{\text{s}}$ and $\rho_{\text{p}}$, such that $P = \sum_{i\in\{\text{s},\text{p}\}} \rho_i + \sum_{i,j\in\{\text{s},\text{p}\}} B_{ij}\rho_i\rho_j + \mathcal{O}(\rho^3)$.
For the Flory--Huggins model, this yields the coefficients
\begin{equation}
  B^{\text{FH}}_{\text{ss}} = \frac{N^2_{\text{s}} v_0}{2}, \hspace{0.75em}
  B^{\text{FH}}_{\text{sp}} = \frac{N_{\text{s}} N_{\text{p}} v_0}{2}, \hspace{0.75em}
  B^{\text{FH}}_{\text{pp}} = \frac{N^2_{\text{p}} v_0}{2},
  \label{eq:B_FH}
\end{equation}
whereas the coefficients derived from our excluded-volume model, \eqsref{eq:mu_EFH_p} and (\ref{eq:mu_EFH_s}), are
\begin{equation}
  B^{\text{ev}}_{\text{ss}} = \frac{N_{\text{s}} v_0}{2}, \hspace{0.75em}
  B^{\text{ev}}_{\text{sp}} = \frac{N_{\text{s}} v_0}{2}, \hspace{0.75em}
  B^{\text{ev}}_{\text{pp}} = \frac{N^2_{\text{p}} v_0}{2}.
  \label{eq:B_EFH}
\end{equation}
The $B_{\text{pp}}$ coefficients are the same in both expressions as expected, since the polymers are treated equivalently in the two models.
However, the second-virial coefficients involving the S molecules differ.
$B^{\text{ev}}_{\text{ss}}$ is a more physically reasonable result than $B^{\text{FH}}_{\text{ss}}$, since the S molecules are assumed to be compact, rigid units.
In fact, $B^{\text{ev}}_{\text{ss}}$ is the correct result for monodisperse hard particles (such as hard spheres~\cite{hansen2013theory}) that exclude a volume of $N_{\text{s}} v_0$.
The cross term, $B^{\text{ev}}_{\text{sp}}$, is less obvious, but we can build intuition by considering the insertion of polymers into a low-density gas of S molecules.
For simplicity, let us assume that the S molecules are hard spheres with diameter $D \simeq N_{\text{s}}^{1/3}v_0^{1/3}$.
The average end-to-end distance of the freely jointed polymer chains is $d \simeq v_0^{1/3}N_{\text{p}}^{1/2}$.
A simple approximation of the cross term is therefore ${B^{\text{ev}}_{\text{sp}} \approx (D + d)^3/2 = N_{\text{s}}(1 + r)^3v_0/2}$, where $r = N_{\text{p}}^{1/2}/N_{\text{s}}^{1/3}$.
Our excluded-volume model of S/P mixtures is therefore most accurate when the S molecules are large compared to the individual polymer chains (such as in the case considered in \secref{sec:nonconserved-volume}), while the standard Flory--Huggins model is more appropriate when the size of an S molecule is comparable to that of an individual polymer blob.
We therefore conclude that the excluded-volume formulae \eqsref{eq:mu_EFH_p} and (\ref{eq:mu_EFH_s}) are the more appropriate choice for the model and parameter regimes considered in this study.

\end{document}